\documentclass{jjap3}
\usepackage{newtxtext,newtxmath}
\usepackage{graphicx}

\title{Robust-fidelity hyperparallel controlled-phase-flip gate through  microcavities}

\author{Hai-Rui Wei$^{1}$\thanks{E-mail: hrwei@ustb.edu.cn}, Yan-Bei Zheng$^{1}$, Ming Hua$^{2}$, and Guo-Fu Xu$^{3}$}

\inst{$^{1}$ School of Mathematics and Physics, University of Science and Technology Beijing, Beijing 100083, China \\

$^{2}$ Department of Applied Physics, Tianjin Polytechnic University, Tianjin 300387, China\\

$^{3}$ Physics Department, Shandong University, Jinan 250100, China}

\abst{Hyperparallel quantum information processing outperforms its traditional parallel one in terms of channel capacity, low loss rate, and processing speed. We present a way for implementing  a robust hyper-parallel optical controlled-phase-flip gate through microcavities.  The gate acts on polarization and spatial degrees of freedom (DOFs) simultaneously, and the incomplete and undesired interactions between photons and quantum dots are prevented. Interestingly, the unity fidelity of the gate can be achieved in principle, and the success of the gate is heralded by the single-photon detectors.}

\begin{document}

\maketitle


Precise manipulations of quantum states are very necessary for quantum information processing (QIP) tasks, and they can be achieved by employing quantum gates. Universal controlled-NOT (CNOT) gates or the equivalent controlled-phase-flip (CPF) gates \cite{universal} are a promising and flourishing area of research filed, that because CNOT (CPF) gates are essential for various QIP protocols. Considerable theoretical and experimental progresses have been made  on CNOT (CPF) gates \cite{KLM,CNOT-cross,Duan,trap-Ion,Long2,atom20141,atom20142,atom2016,linear,CNOT-multiphoton,photon-pnoton-Rydberg,neutral-atom}. Ion-trap CNOTs \cite{trap-Ion}, nuclear magnetic resonance CNOTs \cite{Long2}, photon-atom CNOTs \cite{atom20141,atom20142}, optical CNOTs \cite{atom2016,linear,CNOT-multiphoton,photon-pnoton-Rydberg}, and neutral atom CNOTs  \cite{neutral-atom} have been demonstrated in experiments recently.

Hyperparallel (hyper-) QIP, different from the traditional parallel QIP that is operated on only one  degree of freedom (DOF), is simultaneously performed on multiple qubit-like independent DOFs. Notably, QIP with multiple DOFs has been shown advantages in high capacity, low loss rate, less quantum resources, and loss decoherence characters. For example, different from polarization photons, the frequency, spatial, and time-bin photons are immune to the decoherence effects and the bit-flip errors in optical fibers.  Moreover, hyperentangled states can be used to complete some tasks which are impossible to achieve with the single DOF, such as entanglement witness \cite{witness}, linear-optical Bell-state analysis \cite{Bell-measurement}, and one-way quantum computing \cite{one-way}.
Currently, hyperentangled cat states \cite{cat1}, complete hyperentangled Bell-states and  hyperentangled Greenberger-Horne-Zeilinger states analysis \cite{Analisis1,Analisis3},  hyperentanglement purification \cite{purification2} and concentration \cite{concentration2}, hyper-repeater \cite{Repeater1,Repeater2}, and hyper-transistor \cite{transistor}, and hyper-CNOT and hyper-Toffoli gates \cite{Wei-OE,Litao1,Du-CNOT,Wei-Ann} have been proposed.

Photons have been widely recognized as promising natural candidates for hyper-QIPs owing to their multiple independent qubit-like DOFs, low decoherence, flexible single-qubit manipulation, and high-speed transmission. Unfortunately, scalability is prohibited in one-photon multi-qubit QIP \cite{time-bin}, and interactions between individual photons are necessary for scalable QIPs.  So far, only probabilistic gates can be achieved by using single-photon sources and linear optical elements \cite{KLM}. Interestingly, deterministic multi-photon QIP can be achieved by employing cavity quantum electrodynamics with atoms \cite{Duan} and artificial atoms (such as quantum dot (QD) \cite{QD}, Josephson junction \cite{Josephon}, nitrogen-vacancy  color centers in diamond \cite{NV1,NV2}), or by employing cross-Kerr \cite{CNOT-cross} in principle. Realizing a giant natural cross-Kerr phase shift is still a challenge with current technology \cite{weak-Kerr}. The fast manipulation and readout of neutral atoms are now not established in experiment  \cite{atom-coherence}.

\begin{figure}
\includegraphics[width=9 cm,angle=0]{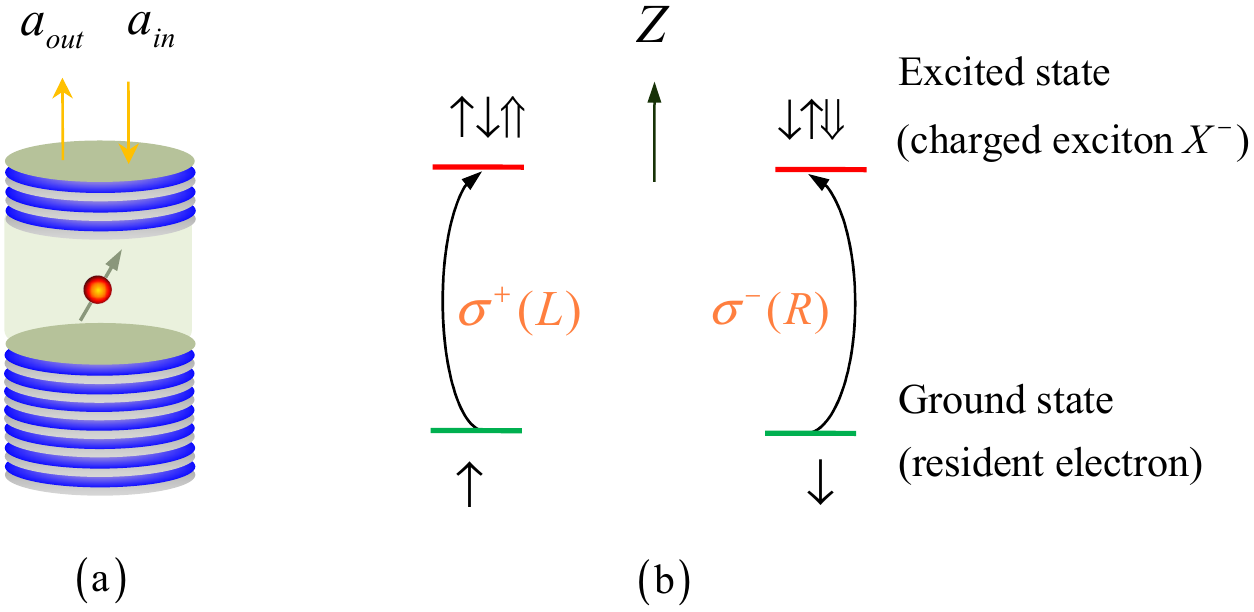}
\caption{The graphic of photon-QD interactions. (a)  A negatively charged exciton $X^{-}$ confined into the single-sided microcavity. (b) The spin-dependence optical transition rules.}
\label{level}
\end{figure}

In this paper, we design a compact quantum circuit for implementing the robust and heralded hyper-CPF gate assisted by QD-cavity systems. A single electron charged GaAs/Al(Ga)As QD located in the center of a micropillar cavity with asymmetric distributed Bragg reflectors \cite{Hu1} is considered. As shown in Fig. \ref{level}, a controllable negatively charged exciton ($X^-$) can be created by injecting an excess electron into the QD. The spin-dependence transitions of the $X^-$ are strictly governed by pauli's exclusion principle and conservation of angular momentum. In detail, the transition $|\uparrow\rangle\rightarrow|\uparrow\downarrow\Uparrow\rangle$ ($|\downarrow\rangle\rightarrow|\downarrow\uparrow\Downarrow\rangle$) is only driven by the left- (right-) handed circularly polarized photon $|L\rangle$ ($|R\rangle$). $|\uparrow\rangle$  and $|\downarrow\rangle$ are the electron-spin states with $J_z=\pm1/2$, respectively.  $|\Uparrow\rangle$ and $|\Downarrow\rangle$ are the heavy-hole states with $J_z=\pm3/2$, respectively. The other transitions are forbidden \cite{forbid}.

According to the quantum mechanic principle,  the interaction processes between cavity-QD systems and circularly polarized photons could be described by Heisenberg equations for the cavity field operator $\hat{a}$ and dipole operator $\hat{\sigma}_-$ \cite{Heisenberg}
\begin{equation}   \label{eq1}
\begin{split}
&\frac{d \hat{a}}{dt}=-\left[i(\omega_c-\omega)+ \frac{\kappa}{2}+\frac{\kappa_s}{2}\right] \hat{a}-g\;\hat{\sigma}_{-} -\sqrt{\kappa}\;\hat{a}_{in}+\hat{R},  \\
&\frac{d\hat{\sigma}_-}{dt}=-\left[i(\omega_{X^-}-\omega)+\frac{\gamma}{2}\right]\hat{\sigma}_{-}-g\;\hat{\sigma}_z\hat{a}+\hat{N},
\end{split}
\end{equation}
where $\omega_{X^-}$, $\omega_{c}$ and $\omega$ are the frequencies of the $X^-$, the cavity mode, and the incident single photon, respectively. $g$ is the $X^-$-cavity coupling strength. $\gamma/2$ and $\kappa/2$ are the decay rates of the $X^-$ and the cavity mode, respectively. $\kappa_{s}/2$ is the side leakage rate of the cavity mode. $\hat{R}$ and $\hat{N}$ are the noise operations related to reservoirs, which can preserve the desired commutation relations.

Considering weak-exciting, and combing the Heisenberg equation and the input-output relation $\hat{a}_\text{out}=\hat{a}_\text{in}+\sqrt{\kappa}\hat{a}$, one can get the reflection coefficient of cavity \cite{Hu1,An-JunHong}
\begin{eqnarray}   \label{eq2}
r(\omega)=1-\frac{\kappa\left[i(\omega_{X^-}-\omega)+\frac{\gamma}{2}\right]}{\left[i(\omega_{X^-}-\omega)+\frac{\gamma}{2}\right]\left[i(\omega_{c}-\omega)+\frac{\kappa}{2}+\frac{\kappa_s}{2}\right]+\;g^2}.
\end{eqnarray}
Such platform can be served as an emitter
\begin{eqnarray}   \label{eq3}
\begin{split}
&|F\rangle|+\rangle \xrightarrow{\text{QD}} \frac{1}{2} ((r_0-r_h)|S\rangle|-\rangle)+(r_0+r_h)|F\rangle|+\rangle),\\
&|F\rangle|-\rangle \xrightarrow{\text{QD}} \frac{1}{2} ((r_0-r_h)|S\rangle|+\rangle+(r_0+r_h)|F\rangle|-\rangle), \\
&|S\rangle|+\rangle \xrightarrow{\text{QD}} \frac{1}{2} ((r_0-r_h)|F\rangle|-\rangle+(r_0+r_h)|S\rangle|+\rangle),\\
&|S\rangle|-\rangle \xrightarrow{\text{QD}} \frac{1}{2} ((r_0-r_h)|F\rangle|+\rangle+(r_0+r_h)|S\rangle|-\rangle).
\end{split}
\end{eqnarray}
Here, $r_0$ and $r_h$ are the reflection coefficients in the cold cavity ($g=0$, i.e., QD does not couple to the cavity) and hot cavity ($g\neq0$, i.e., QD couples to the cavity), respectively. $|F\rangle=\frac{1}{\sqrt{2}}(|R\rangle+|L\rangle)$, $|S\rangle=\frac{1}{\sqrt{2}}(|R\rangle-|L\rangle)$, and $|\pm\rangle=\frac{1}{\sqrt{2}}(|\uparrow\rangle\pm|\downarrow\rangle)$.
However, the incomplete (error) photon-QD interaction is inevitable in experiments. If the incomplete photon-QD interaction with the probability $1-p^2$ is taken into account, Eq. (\ref{eq3}) will be changed as
\begin{eqnarray}   \label{eq4}
\begin{split}
&|F\rangle|+\rangle \xrightarrow{\text{QD}} c|S\rangle|-\rangle + f|F\rangle|+\rangle,\\
&|F\rangle|-\rangle \xrightarrow{\text{QD}} c|S\rangle|+\rangle + f|F\rangle|-\rangle,\\
&|S\rangle|+\rangle \xrightarrow{\text{QD}} c|F\rangle|-\rangle + f|S\rangle|+\rangle,\\
&|S\rangle|-\rangle \xrightarrow{\text{QD}} c|F\rangle|+\rangle + f|S\rangle|-\rangle.
\end{split}
\end{eqnarray}
Here, the complete photon-QD interaction coefficient $p\in(0, 1]$, $c=\frac{p}{2}(r_0-r_h)$ and $f=\frac{p}{2}(r_0+r_h)+\sqrt{1-p^2}$.

\begin{figure}
\includegraphics[width=9.0 cm,angle=0]{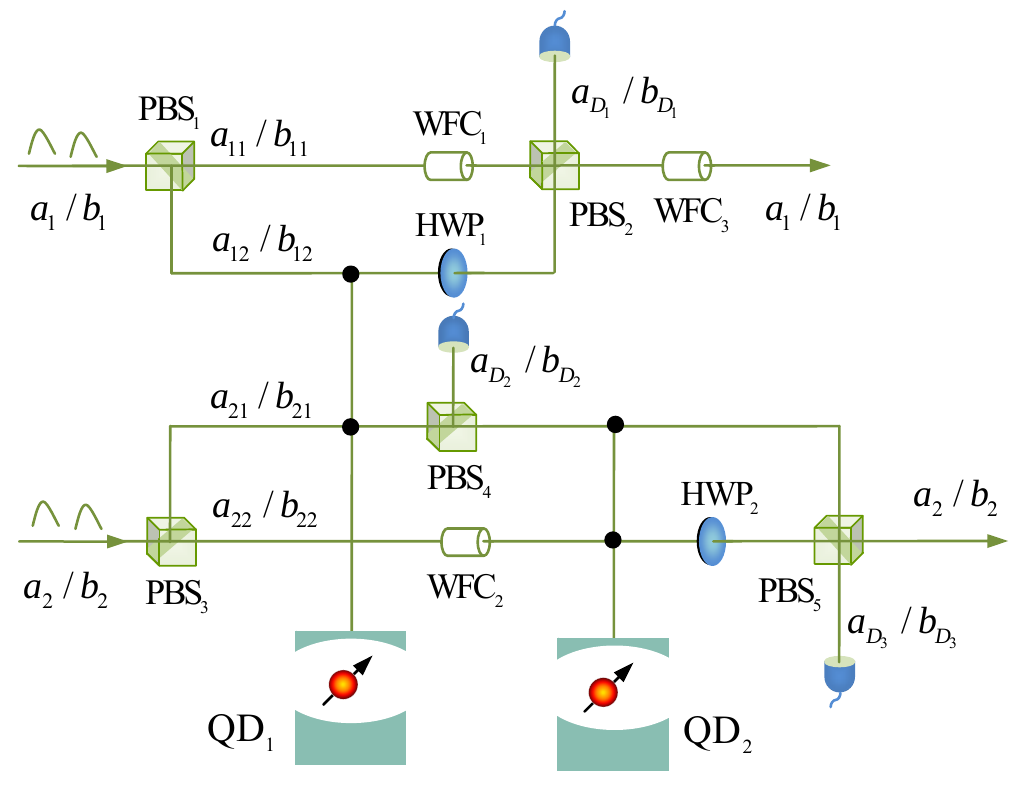}
\caption{A schematic setup for implementing a robust-fidelity hyper-CPF gate.
PBS is a polarization beam splitter oriented in the $\{F,S\}$ basis to transmit the $F$-polarized photon and reflect the $S$-polarized photon, respectively.
HWP is a half-wave plate rotated at $0^{\circ}$ to perform transformation $|F\rangle \stackrel{\text{HWP}}{\longleftrightarrow} |S\rangle$.
Wave-form corrector (WFC) completes $|F\rangle \stackrel{\text{WFC}}{\longrightarrow} c |F\rangle$ and $|S\rangle \stackrel{\text{WFC}}{\longrightarrow} c |S\rangle$ with $c=\frac{p}{2}(r_0-r_h)$. $D_i$ ($i$=1, 2, 3) are single-photon detectors. ``$\bullet$'' indicates a circulator. }
\label{gate}
\end{figure}

Based on the emitter described by Eq. (\ref{eq4}), we propose a protocol for implementing a robust-fidelity hyper-CPF gate (see Fig. \ref{gate}). Notably, the CPF gate and the CNOT gate with control qubit ``$c$'' and target qubit ``$t$'' are equivalent up to the local Hadamard operations performed on the target qubit.  Suppose the control photon $c$, the target photon $t$,  and two solid-state mediates QD$_1$ and QD$_2$ are initially prepared into the normalization states
\begin{eqnarray}   \label{eq5}
\begin{split}
&|\varphi_{c}\rangle=(\alpha_{1}|F\rangle+\alpha_{2}|S\rangle)\otimes(\beta_{1}|a_{1}\rangle+\beta_{2}|a_{2}\rangle),\\
&|\varphi_{t}\rangle=(\lambda_{1}|F\rangle+\lambda_{2}|S\rangle)\otimes(\varpi_{1}|b_{1}\rangle+\varpi_{2}|b_{2}\rangle),\\
&|\varphi_{1}\rangle=|+_1\rangle,\quad
|\varphi_{2}\rangle=|+_2\rangle.
\end{split}
\end{eqnarray}
Here, $|a_1\rangle$ ($|b_1\rangle$) and $|a_2\rangle$ ($|b_2\rangle$) are the two spatial DOFs of the first (second) photons, respectively. $|+_1\rangle$  and $|+_2\rangle$ denote the states of QD$_1$ and QD$_2$, respectively.

Firstly,  as shown in Fig. \ref{gate}, the control photon $c$ in $a_1$ and $a_2$ passes through polarizing beam splitter (PBS$_1$ and PBS$_3$) oriented in the $FS$-basis, respectively. Next, the $F$-polarized components in spatial mode $a_{11}$ ($a_{22}$)  pass through WFC$_1$ (WFC$_2$), and the $S$-polarized components in spatial mode $a_{12}$ ($a_{21}$) interact with QD$_1$ and HWP$_1$ (QD$_1$). Above operations ($\text{PBS}_1 \rightarrow \text{QD}_1 \rightarrow \text{HWP}_1, \text{WFC}_1$ and $\text{PBS}_3 \rightarrow \text{QD}_1, \text{WFC}_2$) transform the composite system from  $|\psi_0\rangle=|\varphi_{c}\rangle\otimes|\varphi_{t}\rangle\otimes|\varphi_{1}\rangle\otimes|\varphi_{2}\rangle$ into
\begin{eqnarray}   \label{eq6}
\begin{split}
|\psi_1\rangle=&
(c\alpha_{1}\beta_{1}|F\rangle|a_{11}\rangle|+_1\rangle + c\alpha_{1}\beta_{2}|F\rangle|a_{22}\rangle|+_1\rangle\\&
+c\alpha_{2}\beta_{1}|S\rangle|a_{12}\rangle|-_1\rangle + f\alpha_{2}\beta_{1}|F\rangle|a_{12}\rangle|+_1\rangle\\&
+c\alpha_{2}\beta_{2}|F\rangle|a_{21}\rangle|-_1\rangle + f\alpha_{2}\beta_{2}|S\rangle|a_{21}\rangle|+_1\rangle) \otimes |\varphi_{t}\rangle\otimes|\varphi_{2}\rangle.
\end{split}
\end{eqnarray}
Here and henceforth, half-wave plate, HWP, is rotated at $0^{\circ}$ to perform the qubit-flip transformation $|F\rangle \stackrel{\text{HWP}}{\longleftrightarrow} |S\rangle$. Wave-form corrector (WFC) completes the operations $|F\rangle \xrightarrow{\text{WFC}}  c |F\rangle$ and $|S\rangle \xrightarrow{\text{WFC}}  c |S\rangle$, and it can be achieved by employing an unbalanced beam splitter \cite{BS} or employing  quarter-wave plates, HWPs, and PBSs \cite{Ren-split}.

Secondly, the wave-packets emitted from spatial modes $a_{11}$ and $a_{12}$ mix at PBS$_2$ and then are followed by WFC$_3$.
The wave-packets emitted from $a_{21}$ are split by PBS$_4$, and then undergo QD$_2$ and PBS$_5$ in succession.
The wave-packets emitted from $a_{22}$ interact with QD$_2$, HWP$_2$ and PBS$_5$ in succession. These operations induce $|\psi_1\rangle$ to become
\begin{eqnarray}   \label{eq7}
\begin{split}
|\psi_2\rangle=&
(cc \alpha_{1}\beta_{1}|F\rangle|a_{1}\rangle|+_1\rangle|+_2\rangle
+cc\alpha_{1}\beta_{2}|F\rangle|a_{2}\rangle|+_1\rangle|-_2\rangle\\&
+
cf\alpha_{1}\beta_{2}|S\rangle|a_{D_3}\rangle|+_1\rangle|+_2\rangle
+cc\alpha_{2}\beta_{1}|S\rangle|a_{1}\rangle|-_1\rangle|+_2\rangle\\&
+f\alpha_{2}\beta_{1}|F\rangle|a_{D_1}\rangle|+_1\rangle|+_2\rangle
+cc\alpha_{2}\beta_{2}|S\rangle|a_{2}\rangle|-_1\rangle|-_2\rangle\\&
+cf\alpha_{2}\beta_{2}|F\rangle|a_{D_3}\rangle|-_1\rangle|+_2\rangle
+f\alpha_{2}\beta_{2}|S\rangle|a_{D_2}\rangle|+_1\rangle|+_2\rangle) \otimes |\varphi_{t}\rangle.
\end{split}
\end{eqnarray}
Based on Eqs. (\ref{eq4}-\ref{eq7}) and Fig. \ref{gate}, we can see that the incomplete photon-QD interaction components $|F\rangle|a_{D_1}\rangle$, $|S\rangle|a_{D_3}\rangle$, $|F\rangle|a_{D_3}\rangle$ and $|S\rangle|a_{D_2}\rangle$ are prevented (detected) by the single-photon detectors.

Thirdly, we apply Hadamard operations $H_e$s on QD$_1$ and QD$_2$, which can be characterized by
\begin{eqnarray}   \label{eq8}
|\uparrow\rangle \stackrel{H_e}{\longleftrightarrow} |+\rangle=\frac{1}{\sqrt{2}}(|\uparrow\rangle+|\downarrow\rangle),\qquad
|\downarrow\rangle \stackrel{H_e}{\longleftrightarrow} |-\rangle=\frac{1}{\sqrt{2}}(|\uparrow\rangle-|\downarrow\rangle).
\end{eqnarray}

Next, the target photon $t$ is injected. After similar arguments as that made for the first photon, when all the single-photon detectors are not clicked, the system collapses into the state
\begin{eqnarray}   \label{eq9}
\begin{split}
|\psi_3\rangle=&
 c^4\alpha_{1} \beta_{1}|F\rangle|a_{1}\rangle(\lambda_1|F\rangle+\lambda_2|S\rangle)(\varpi_1|b_1\rangle+\varpi_2|b_2\rangle)|\uparrow_1\uparrow_2\rangle\\ &
+c^4\alpha_{1}\beta_{2}|F\rangle|a_{2}\rangle(\lambda_1|F\rangle+\lambda_2|S\rangle)(\varpi_1|b_1\rangle-\varpi_2|b_2\rangle)|\uparrow_1\downarrow_2\rangle\\ &
+c^4\alpha_{2}\beta_{1}|S\rangle|a_{1}\rangle(\lambda_1|F\rangle-\lambda_2|S\rangle)(\varpi_1|b_1\rangle+\varpi_2|b_2\rangle)|\downarrow_1\uparrow_2\rangle\\ &
+c^4\alpha_{2}\beta_{2}|S\rangle|a_{2}\rangle(\lambda_1|F\rangle-\lambda_2|S\rangle)(\varpi_1|b_1\rangle-\varpi_2|b_2\rangle)|\downarrow_1\downarrow_2\rangle.\\
\end{split}
\end{eqnarray}

Finally, the spins of the QD$_1$ and QD$_2$ are measured in the $\{|\pm\rangle\}$ basis, and after some classical feed-forward operations (see Table \ref{table}) are performed on the first photon, the state of the whole system becomes
\begin{eqnarray}   \label{eq13}
\begin{split}
|\psi_4\rangle=&c^4 (\alpha_{1}\lambda_{1}|F\rangle|F\rangle + \alpha_{1}\lambda_{2}|F\rangle|S\rangle
                   + \alpha_{2}\lambda_{1}|S\rangle|F\rangle - \alpha_{2}\lambda_{2}|S\rangle|S\rangle)\\&
           \otimes(\beta_{1}\varpi_{1}|a_{1}\rangle|b_{1}\rangle + \beta_{1}\varpi_{2}|a_{1}\rangle|b_{2}\rangle
                 + \beta_{2}\varpi_{1}|a_{2}\rangle|b_{1}\rangle - \beta_{2}\varpi_{2}|a_{2}\rangle|b_{2}\rangle).
\end{split}
\end{eqnarray}


Note that the performance of the gate is characterized by the fidelity $\mathcal{F}=|\langle \varphi_{\text{real}}| \varphi_{\text{ideal}}\rangle|$, where  $\varphi_{\text{real}}$ and $\varphi_{\text{ideal}}$ are the normalization outcome  states of the gate in realistic and ideal environments, respectively.
From Eq. (\ref{eq13}), we see that the quantum circuit shown in Fig. \ref{gate}  implemented a unity-fidelity hyper-CPF gate which performs CPF operations on the polarization and the spatial DOFs, simultaneously. The incomplete QD-photon interactions and the undesired QD-cavity interactions are all prevented.

\begin{table}
\caption{The classical feed-forward operations applied on the outing photons to complete the hyper-CPF gate. Polarized-$\sigma_z$ (spatial-$\sigma_z$) can be achieved by employing HWP rotated to 45$^\circ$ (phase shifter $e^{i\pi}$).}
\label{table}
\begin{tabular}{llllll}
\Hline
\multicolumn {2}{c}{Measurement\quad}     & \multicolumn {2}{c}{Feed-forward} \\
\Hline
     QD$_1$    &  QD$_2$          & the photon $c$                       & the photon $t$  \\
\Hline
$|+_1\rangle$  &  $|+_2\rangle$   &   none                                    &   none \\ 

$|+_1\rangle$  &  $|-_2\rangle$   & spatial-$\sigma_z$                        &   none       \\  

$|-_1\rangle$  &  $|+_2\rangle$   & polarized-$\sigma_z$                      &   none  \\ 

$|-_1\rangle$  &  $|-_2\rangle$   & spatial-$\sigma_z$, polarized-$\sigma_z$  &   none \\  
\Hline
\end{tabular}
\end{table}

\begin{figure}
\includegraphics[width=7.0 cm,angle=0]{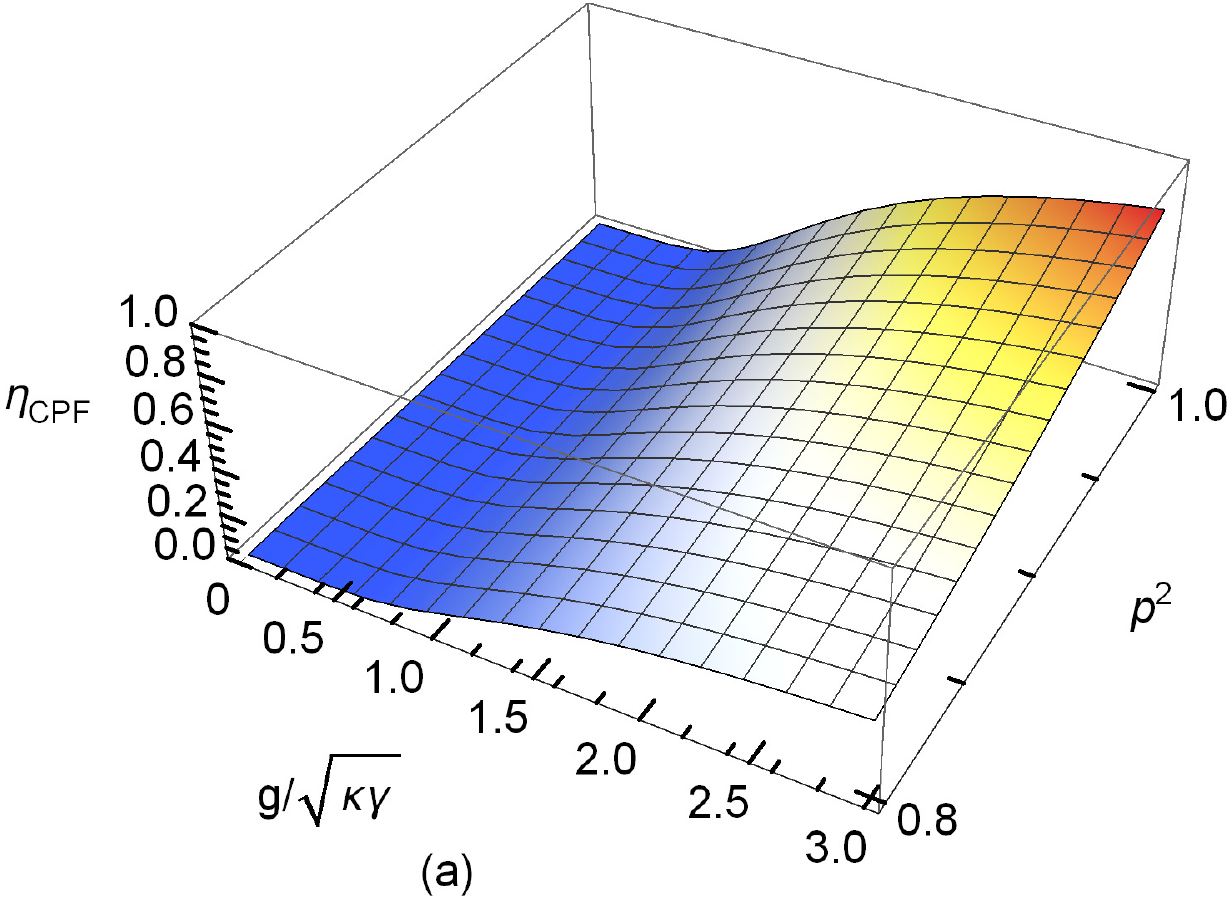}
\qquad \includegraphics[width=7.0 cm,angle=0]{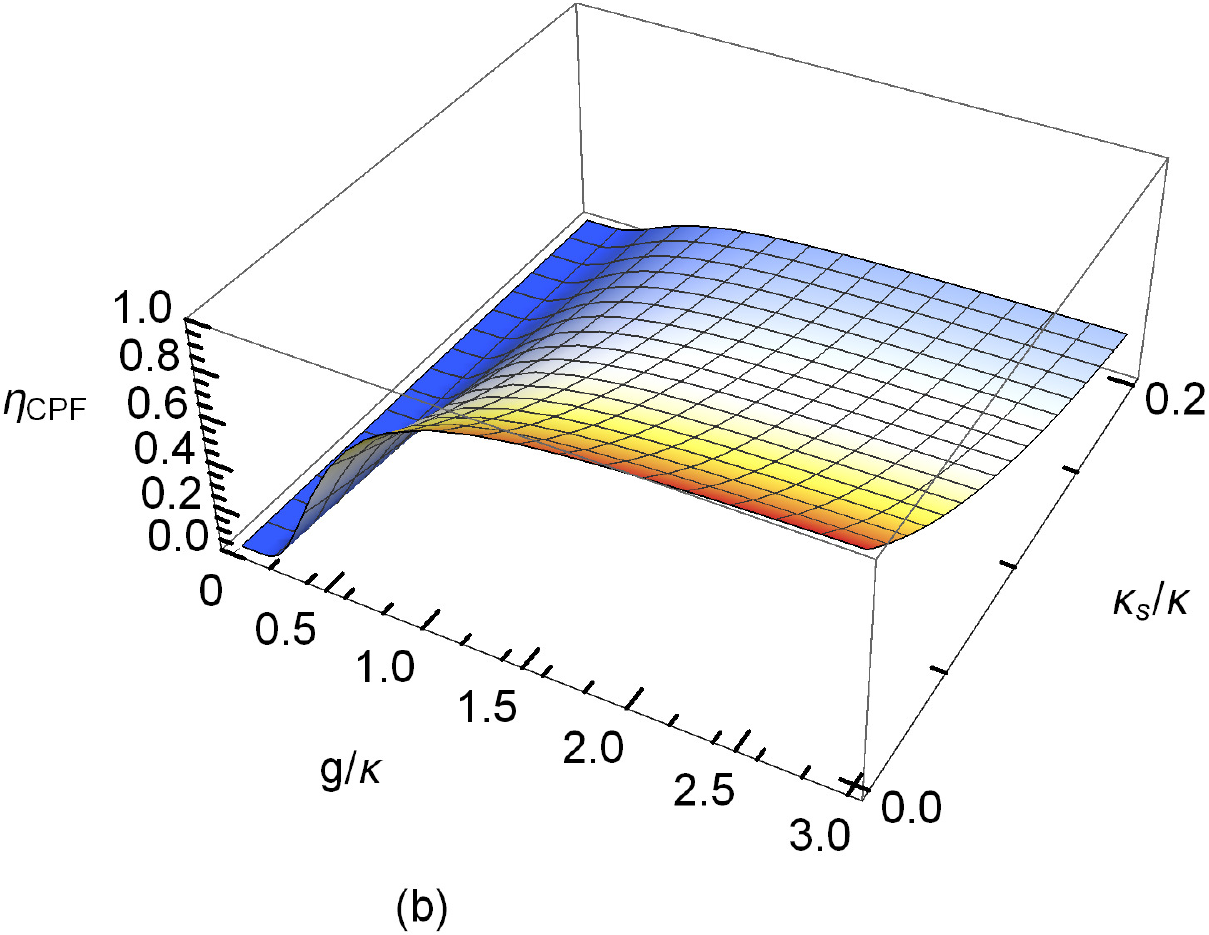}
\caption{(a) The efficiency of the robust-fidelity hyper-CPF gate based on experimental value $\kappa_s\ll\kappa$ \cite{sideleakage}. (b) The efficiency of the robust-fidelity hyper-CPF gate with $p=1$ and experimental value $\gamma=0.1\kappa$.  $\omega=\omega_c=\omega_{X^-}$ is taken for (a) and (b).}
\label{Efficiency}
\end{figure}

Next, we evaluate the efficiency $\mathcal{\eta}=n_{\text{out}}/n_{\text{in}}$ of the present scheme, where $n_{\text{out}}$ and $n_{\text{in}}$  are the number of the output and  input photons, respectively. Based on Eq. (\ref{eq13}),  the efficiency of the present gate can be written as
\begin{eqnarray}   \label{eq14}
\mathcal{\eta}=c^8=\frac{1}{256}p^8(r_0-r_h)^8.
\end{eqnarray}
If $\omega=\omega_c=\omega_{X^-}$ and $\gamma=0.1\kappa$ are taken in the practical QD-cavity parameters, Eq. (\ref{eq14}) yields
\begin{eqnarray}   \label{eq15}
\mathcal{\eta}=\frac{65536g^{16}p^8\kappa^8}{(\kappa_s+\kappa)^8(4g^2+\frac{1}{10}\kappa(\kappa_s+\kappa))^8}.
\end{eqnarray}
Figure \ref{Efficiency}(a) and \ref{Efficiency}(b) depict the efficiencies of the designed hyper-CPF gate as functions of $g/\sqrt{\kappa\gamma}$ and $p^2$, $g/\kappa$ and $\kappa_s/\kappa$, respectively. In addition, the spatial mismatches between the cavity and the incident photon, the qualities of QD spin state preparation, manipulation, and readout, the imperfections in linear optical elements reduce the performance of our proposal. The pillar microcavity with $g/{2\kappa+\kappa_s}=2.4$, $g=80$ $\mu$eV, $2\kappa+\kappa_s=33 $ $\mu$eV, and $Q=7\times 10^4$  has been experimentally achieved \cite{sideleakage}. Besides, $g=16$ $\mu$eV, $\kappa=20.5$ $\mu$eV, and $Q=65000$ have been
reported in micropillar with diameter $d=7.3$ $ \mu$m, recently \cite{g=16}.

In summary, by encoding the gate qubits in the polarization and the spatial states of single-photons, we have proposed a robust  hyper-parallel CPF gate with the help of the single-trapped QD mediates. In the optical architecture, two individual single-photons are bridged by QD, and the inevitable incomplete QD-photon interactions are prevented by single-photon detectors. Interestingly, the unity-fidelity and high efficiency can be achieved in principle, and the success of the gate is heralded by the single-photon detectors. The strong coupling limitations and the auxiliary single-photons are not necessary in our scheme. The proposed proposal is possible with the present technology and it can be used in robust hyper-parallel quantum computing and quantum communication.

\acknowledgment

The work is supported by the National Natural Science Foundation of China under Grant No. 11604012, and the Fundamental Research Funds for the Central Universities under Grant Nos.  FRF-BR-17-004B and 230201506500024, and a grant from China Scholarship Council.

\end{document}